# Proximity-Induced Superconductivity in Epitaxial Topological Insulator/Graphene/Gallium Heterostructures

Cequn Li[1], Yi-Fan Zhao[1], Alexander Vera[2,3], Omri Lesser[4], Hemian Yi[1], Shalini Kumari[2,3], Zijie Yan[1], Chengye Dong[5], Timothy Bowen[2,3], Ke Wang[6], Haiying Wang[6], Jessica L. Thompson[7], Kenji Watanabe[8], Takashi Taniguchi[9], Danielle Reifsnyder Hickey[3,6,7], Yuval Oreg[4], Joshua A. Robinson[1,2,3,5,6,7], Cui-Zu Chang[1], and Jun Zhu[1,2,5*]

1. Department of Physics, The Pennsylvania State University, University Park, PA 16802, USA
2. Center for 2-Dimensional and Layered Materials, The Pennsylvania State University, University Park, PA 16802, USA
3. Department of Materials Science and Engineering, The Pennsylvania State University, University Park, PA 16802, USA
4. Department of Condensed Matter Physics, Weizmann Institute of Science, Rehovot 760001, Israel
5. 2-Dimensional Crystal Consortium, The Pennsylvania State University, University Park, PA 16802, USA.
6. Materials Research Institute, The Pennsylvania State University, University Park, PA 16802, USA
7. Department of Chemistry, The Pennsylvania State University, University Park, PA 16802, USA
8. Research Center for Functional Materials, National Institute for Materials Science, 1-1 Namiki, Tsukuba 305-0044, Japan
9. International Center for Materials Nanoarchitectonics, National Institute for Materials Science, 1-1 Namiki, Tsukuba 305-0044, Japan.

* Correspondence to: jzhu@phys.psu.edu (J. Zhu)

**Abstract:** The introduction of superconductivity to the Dirac surface states of a topological insulator leads to a topological superconductor, which may support topological quantum computing through Majorana zero modes[1, 2]. The development of a scalable material platform is key to the realization of topological quantum computing[3, 4]. Here we report on the growth and properties of high-quality (Bi,Sb)$_2$Te$_3$/graphene/gallium heterostructures. Our synthetic approach enables atomically sharp layers at both hetero-interfaces, which in turn promotes proximity-induced superconductivity that originates in the gallium film. A

**lithography-free, van der Waals tunnel junction is developed to perform transport tunneling spectroscopy. We find a robust, proximity-induced superconducting gap formed in the Dirac surface states in 5-10 quintuple-layer $(Bi,Sb)_2Te_3$/graphene/gallium heterostructures. The presence of a single Abrikosov vortex, where the Majorana zero modes are expected to reside, manifests in discrete conductance changes. The present material platform opens up opportunities for understanding and harnessing the application potential of topological superconductivity.**

**Main text:** A primary approach to realize topological superconductivity in a potentially scalable material platform is to exploit the superconducting proximity effect of a hybrid system involving an s-wave superconductor (SC), and a one-dimensional (1D) or two-dimensional (2D) semiconductor with strong spin-orbit coupling[3-5]. Studies following this approach have made much progress in fundamental science and materials engineering, but have also encountered great challenges[4]. As envisioned by Fu and Kane[1], topological superconductivity is also predicted to occur in a hybrid system of a topological insulator (TI) and SC, where the lifting of the spin degeneracy, a key ingredient for topological superconductivity, is already achieved in the helical surface states of a topological insulator. Scanning tunneling microscopy studies have reported proximity-induced superconductivity and signatures of Majorana zero modes in TI/SC heterostructures[6-9]. However, the difficulty of creating high-quality, scalable platforms remains a major roadblock to further experimentation. A high-quality TI film can be grown on bulk $NbSe_2$[6-8], but does not form a clean hetero-junction on metallic film superconductors such as Al or Nb due to interfacial reactions[10]. In addition, to prevent the oxidization of TI films is a challenge in device fabrication. Transport studies, a major experimental probe of Majorana zero modes, remain limited in TI/SC heterostructures[11-14].

In this work, we demonstrate the synthesis of epitaxial $(Bi,Sb)_2Te_3$/graphene/Gallium (BST/Gr/Ga) thin films with atomically sharp hetero-interfaces. We construct clean van der Waals (vdW) tunnel junctions and employ transport tunneling spectroscopy to probe their superconducting properties. Tunneling spectra obtained on heterostructures with the BST thickness ranging from 5 to 10 quintuple layers (QLs) exhibit the characteristics of two superconducting gaps. The larger gap originates from the superconducting Ga film[15]. We attribute the smaller gap, which is 30-50% of the gap in Ga, to proximity-induced superconductivity in the Dirac surface states of the BST film. The introduction of magnetic vortices leads to discrete tunneling conductance jumps, with the smallest unit corresponding to the increment of a single flux quantum. Our experiments established a large area, potentially scalable TI/SC platform for the studies of topological superconductivity and Majorana physics.

As illustrated in Fig. 1a, BST/Gr/Ga heterostructures are synthesized in two steps. We intercalate an atomically thin 2D Ga film at the epi-graphene/6H-SiC (0001) substrates using confinement heteroepitaxy (CHet)[15]. The Ga film covers 90% of the wafer, grows uniformly and epitaxially across the majority of the terraces of the SiC substrate, and consists of mainly two-atomic-layer-thick (2L) Ga, although 1L and 3L-Ga are also present in some regions of the wafer[16, 17]. The 2D Ga films exhibit superconductivity with a transition temperature of approximately 3-4 K, higher than that of the bulk $\alpha$-Ga[18], which is attributed to electron-phonon interactions and the free-electron-like pockets near the Ga K points. A full characterization and understanding of the growth and superconductivity of 2D Ga is given in Ref. [15-17].

We then grow $(Bi,Sb)_2Te_3$ of controlled thickness and composition using molecular beam epitaxy (MBE)[19]. Crucially, the epi-graphene layer acts as a chemical barrier to prevent the formation of Ga-(Bi,Sb) alloys and a Ga-Te compound and templates the growth of BST due to their similar lattice structures[20, 21]. This approach results in abrupt, atomically sharp growth at every hetero-interface, as verified by cross-sectional scanning transmission electron microscopy (STEM; Fig. 1b). In addition, the strain effect induced by lattice mismatches between BST, epi-graphene and Ga is negligible, as the epi-graphene is quasi-freestanding upon Ga intercalation[15], and the strain in MBE-grown $Bi_2Te_3$ family films is fully relaxed on graphene substrate[20]. The clean, vdW-like interface facilitates high-efficiency superconducting proximity coupling[22]. The grown film is transferred from the MBE chamber to an Ar-filled glovebox without breaking vacuum. We construct a clean vdW tunnel junction consisting of a graphite electrode and a 1-2 layer hexagonal boron nitride (h-BN) tunnel barrier[24] (Fig. 1d) and transfer the tunnel junction to the BST/Gr/Ga film inside the glovebox (Fig. 1c). The active area is encapsulated by a large h-BN sheet to prevent oxidization before the film is removed from the glovebox and processed into a full device using lithography (Fig. 1e). Fabrication details are given in Section S2 of the Supplementary Information (SI). The use of a vdW tunnel junction preserves the pristine BST surface and enables measurements under a wide range of experimental conditions outside the growth chamber.

In-situ angle-resolved photoemission spectroscopy (ARPES) measurements are performed on grown BST films to examine the film quality and Fermi level placement $E_F$. Figure 2a shows an exemplary band map of a 5 QL $(Bi_{0.7}Sb_{0.3})_2Te_3$/Gr/Ga film (See Fig. S1 for band maps on 8 and 10QL films). The Dirac surface states (DSS) are sharp and are the only states crossing the Fermi level, which fulfils the prerequisite for the formation of topological superconductivity. The ARPES band structure of films grown with the same chemical composition and thickness reproduces very well in different trials. The Dirac point is located at energy $(E-E_F)$ = -205 meV, which is roughly where the valence band maximum is. Guided by the ARPES results, we associate the occurrence of prominent slope changes on $dI/dV$ spectra ($I$, current; $V$, voltage) taken on the same film with the onset of the bulk conduction and valence bands, following prior studies in the literature[25, 26]. The gray box in Fig. 2b corresponds roughly to the energy range of the bulk gap, consistent with the ARPES band map shown in Fig. 2a.

Owing to the epi-graphene reaction barrier, the growth of BST preserves the superconductivity of the 2L Ga film[15] extremely well. We perform temperature- and magnetic field-dependent transport measurements on a series of heterostructures with varying BST film thickness. In these Hall bar devices, the metal electrodes contact the Gr/Ga film directly through occasional voids of the BST film (See Fig. S3b of the SI for cross-sectional STEM images of the electrode). Our results consistently show a critical temperature $T_c$ of 3–4.2 K and an upper critical perpendicular field $H_{c2}$ of 50–80 mT for the Ga film. The variation of $T_c$ among samples may be caused by the different number of Ga layers and the presence of superconducting weak links in the film. Figure 2c plots an example of longitudinal resistance $R_{xx}$ versus temperature $T$ measured on device 04H-5QL at selected fields. Here 04 is the film number as listed in Table S1 of the SI, H stands for Hall bar, and 5QL is the BST film thickness. Other devices are named following the same scheme. Additional data are given in Section S3 of the SI. Figure 2d plots $H_{c2}$ versus $T$ extracted from Fig.

S3f. The linear relationship between $H_{c2}$ and $T$ confirms the 2D nature of the superconductivity[27]. Our heterostructure film can support a critical current density $J_c$ as large as 0.14 A/μm² (Fig. S3g), which is compared to other elemental superconductor films such as Al and Nb[28].

Next, we perform tunneling spectroscopy measurements on 5QL and 10 QL BST/Gr/Ga films using devices similar to that shown in Fig. 1e. The construction and interfacial structure of the tunnel junction are described in detail in Fig. S2 of the Supplementary Information. The tunneling process is schematically illustrated in Fig. 3a. The tunneling current can either tunnel into the surface states of the BST film then flow to the drain electrode through a distributed network as shown in Fig. 3a, or through the bulk to the Ga layer underneath; this allows us to probe the superconductivity of both layers. We have verified the efficacy of the vdW junction by first applying it to a Gr/Ga film and the resulting $dI/dV(V_{dc})$ spectra at are shown in Fig. 3b. Both the V-shaped dip near zero bias and the accompanying peaks at finite biases are characteristics of a superconducting gap, as observed in scanning tunneling spectroscopy[6, 7, 9] or point contact spectroscopy[11]. Indeed, the tunneling spectra in Fig. 3b can be well fit by a Blonder-Tinkham-Klapwijk (BTK) model[29], from which we obtain the temperature-dependent gap $\Delta(T)$, the tunnel barrier transparency $Z$ and the lifetime broadening $\Gamma$ of the Cooper pairs. (See Section S4 of the SI for details of the BTK model and fits to data.) $\Delta(T)$ is well described by the Bardeen–Cooper–Schrieffer (BCS) theory, i.e., $\Delta(T) = \Delta_0 \tanh(1.74\sqrt{\frac{T_c}{T} - 1})$ (Eq. 1), where $T_c = 3.1$ K is from transport studies of the same film and we obtain the zero-temperature gap $\Delta_0 = 0.395$ meV from the fits (Fig. S4). The parameters of the Gr/Ga film serve as important references to the subsequent studies of BST/Gr/Ga films.

Figure 3c plots the $T$-dependent $dI/dV(V_{dc})$ obtained on device 02T-5QL, which is made on a 5QL BST/Gr/Ga film. We observe the familiar V-shaped dip near zero d.c. bias and not one, but two high conductance humps at high biases. Both the dip and the humps weaken with increasing temperature and disappear completely above 4 K, which corresponds to the $T_c$ of this film obtained in transport studies (Fig. S3d). Not all features of the tunneling spectra are related to superconductivity. We use the magnetic field dependence of the spectra (See Fig. S5b of the SI for the junction shown in Fig. 3c) to identify accidental features coming from the background of the tunnel junction itself, for example, the sharp peak at $V_{dc} = -0.6$ mV in Fig. 3c. Following literature[30], we model the spectra as a weighted sum of two SC tunneling processes. Exemplary fits to data using a two-gap BTK model are shown in Fig. 3d, from which we obtain and label the two gaps as $\Delta_S = 0.19$ meV and $\Delta_I = 0.5$ meV respectively. Similar fits are obtained at other temperatures and the resulting $\Delta_S(T)$ and $\Delta_I(T)$ are plotted in Fig. 3e. Both $\Delta_S(T)$ and $\Delta_I(T)$ are well described by Eq. (1) with a common $T_c$ of 4.0 K. These analyses strongly suggest that $\Delta_I$ originates from the SC gap of the Ga film while $\Delta_S$ is induced by the proximity effect.

Following previous scanning tunneling microscopy studies on TI/SC heterostructures[7, 9], we attribute $\Delta_S$ to the induced gap at the top surface of the BST film, where the Dirac surface states reside, although future studies are needed to further clarify its origin. Single electron tunneling into the BST film onsets at $e|V_{dc}| \geq \Delta_S$, where the top surface turns normal; this gives rise to the zero-bias conductance dip and the first set of coherence peaks. At $e|V_{dc}| \geq \Delta_I$, a second tunneling path opens, where an electron can reach the drain electrode through the bulk of the BST film then the normal Ga film. This leads to the second set of coherence peaks. The fits in Fig. 3d illustrate the individual and the sum of the two tunneling processes. The zero-bias conductance dip and the first coherence peaks are almost entirely given by $\Delta_S$, making its value a very robust finding of the analysis. A proximity-induced SC gap is observed on four tunnel junctions made on film 02T-5QL (Fig. 3f) and our analysis gives $\Delta_S/\Delta_I$ in the range of 37±4% to 46±5% (Table S2 of the SI), which is consistent with previous studies performed on Bi$_2$Se$_3$/NbSe$_2$[6, 11], Bi$_2$Te$_3$/NbSe$_2$[7] and Bi$_2$Se$_3$/Nb[31].

Similar measurements and analysis are performed on two tunnel junctions constructed on 03T-10QL BST/Gr/Ga films, producing $\Delta_S/\Delta_I$ in the range of 34±4% to 42±2%. These measurements attest to the robustness of the proximity-induced superconductivity in the Dirac surface states of the TI film. A more complete and detailed analysis of the tunneling spectra on 5 and 10QL devices are given in Section S5 of the SI. In our setup, tunneling occurs over an area of approximately 3 µm², where the thickness of the BST film could deviate from its nominal value by 1-2 QL. A more precise study of the dependence of $\Delta_S$ on film thickness[6, 7, 11, 31] will require a different design.

In the remainder of the paper, we show tunneling conductance change due to the presence of Abrikosov vortices, which localize Majorana zero modes in a topological superconductor[1, 8]. A vortex is normal at its core and its presence in the tunneling area $A$ raises the zero-bias tunneling conductance by a certain amount $G_{sv}$ (the upper inset of Fig. 4). If the number of vortices in the tunnel junction increases one by one by ramping an out-of-plane magnetic field $B_\perp$, we expect the zero-bias conductance $G = dI/dV$ to increase from its zero-field value in a staircase pattern with an average slope of $\Delta G/\Delta B_\perp = \frac{A \times G_{sv}}{\Phi_0}$, (Eq. 2), where $\Phi_0 = h/2e$ is the magnetic flux quantum and $h$ is the Planck constant. Figure 4 plots the measured zero-bias $dI/dV$ versus $B_\perp$ on device 03T-10QL. While the general trend of the data agrees with Eq. (2), the change of conductance with

$B_\perp$ is not always monotonic and the two field sweep directions show considerable hysteresis, suggesting the trapping of vortices. Furthermore, we observe conductance jumps in multiple integers of $G_{sv}$, suggesting that multiple vortices move together as a bundle[32]. Similar conductance jumps were observed in the transport tunneling spectra on NbSe2 when magnetic vortices appear or disappear from the tunneling area[33]. The lower inset of Fig. 4 show examples of $\Delta G = 1, 2\ G_{sv}$. The above characteristics are general to both Gr/Ga and BST/Gr/Ga junctions, where additional data are given in Section S6 of the SI. These measurements show that transport tunneling spectroscopy provides a valuable probe of the vortex dynamics, which can be performed concomitantly with current-phase studies in Josephson junctions to examine topological phase transitions and Majorana zero modes. These directions will be explored in future studies.

In summary, we demonstrate the growth and characterization of (Bi,Sb)2Te3/Gr/Ga thin-film heterostructures, a potentially scalable platform that could be exploited to realize topological superconductivity. Using a clean vdW tunnel junction and transport tunneling spectroscopy, we show experimental evidence of a robust proximity-induced superconducting gap in the Dirac surface states of the MBE-grown (Bi,Sb)2Te3 film and detect the presence of Abrikosov vortices in tunneling conductance down to a single vortex. Our synthetic approach of combining confinement heteroepitaxy and MBE opens the door to the studies of novel superconducting and magnetic heterostructures. More sophisticated experiments and continued effort to improve the quality and uniformity of the heterostructure film is needed to advance the fundamental understanding and potential applications of topological superconductivity.

**Acknowledgements**

C.L., H.Y., C.-Z.C. S.K., T.B., J.L.T, J.A.R., D.R.H, and J.Z. are supported by the Penn State MRSEC for Nanoscale Science (DMR-2011839). Y.-F.Z, Z.Y. and C.-Z.C. are supported by the NSF-CAREER award (DMR-1847811) and Gordon and Betty Moore Foundation's EPiQS Initiative (GBMF9063 to C. Z. C.). C.D. and J.A.R. are supported by the Penn State NSF-MIP Two-Dimensional Crystal Consortium award (DMR-1539916). A.V. and J.A.R are supported by NSF-DMR 2002651. O.L. and Y.O. are supported by the BSF and NSF (2018643), the European Union's Horizon 2020 research and innovation programme (Grant Agreement LEGOTOP No. 788715), the DFG (CRC/Transregio 183, EI 519/7-1), and ISF Quantum Science and Technology (2074/19). K. Watanabe and T.T. acknowledge support from JSPS KAKENHI (Grant Numbers 19H05790, 20H00354 and 21H05233). J.L.T. and D.R.H. thank the Penn State Eberly College of Science, Department of Chemistry, and Materials Research Institute for generous support through startup funds. The co-authors acknowledge use of the Penn State Materials Characterization Lab. We thank Chao-Xing Liu, Ruobing Mei, Ying Liu for helpful discussions and Ruoxi Zhang and Furkan Turker for assistance in measurement.

**Author contributions:**

C.L. and J.Z. designed the experiment. C.L. fabricated the devices and made the transport measurements under the supervision of J.Z.; Y.-F.Z., H.Y. and Z.Y. performed the MBE growth and ARPES measurements under the supervision of C.-Z.C.; A.V., S.K., C.D. and T.B. performed the CHet growth and characterizations under the supervision of J.A.R; O.L. performed BTK modeling under the supervision of Y.O.; K. Watanabe and T.T. synthesized the h-BN crystals; K. Wang, H.W., and J.L.T. under the supervision of D.R.H performed FIB and TEM measurements; C.L. and J.Z. analyzed data; C.L. and J.Z. wrote the manuscript with input from all authors.

**Competing interests:**

The authors declare no competing interests.

Figures

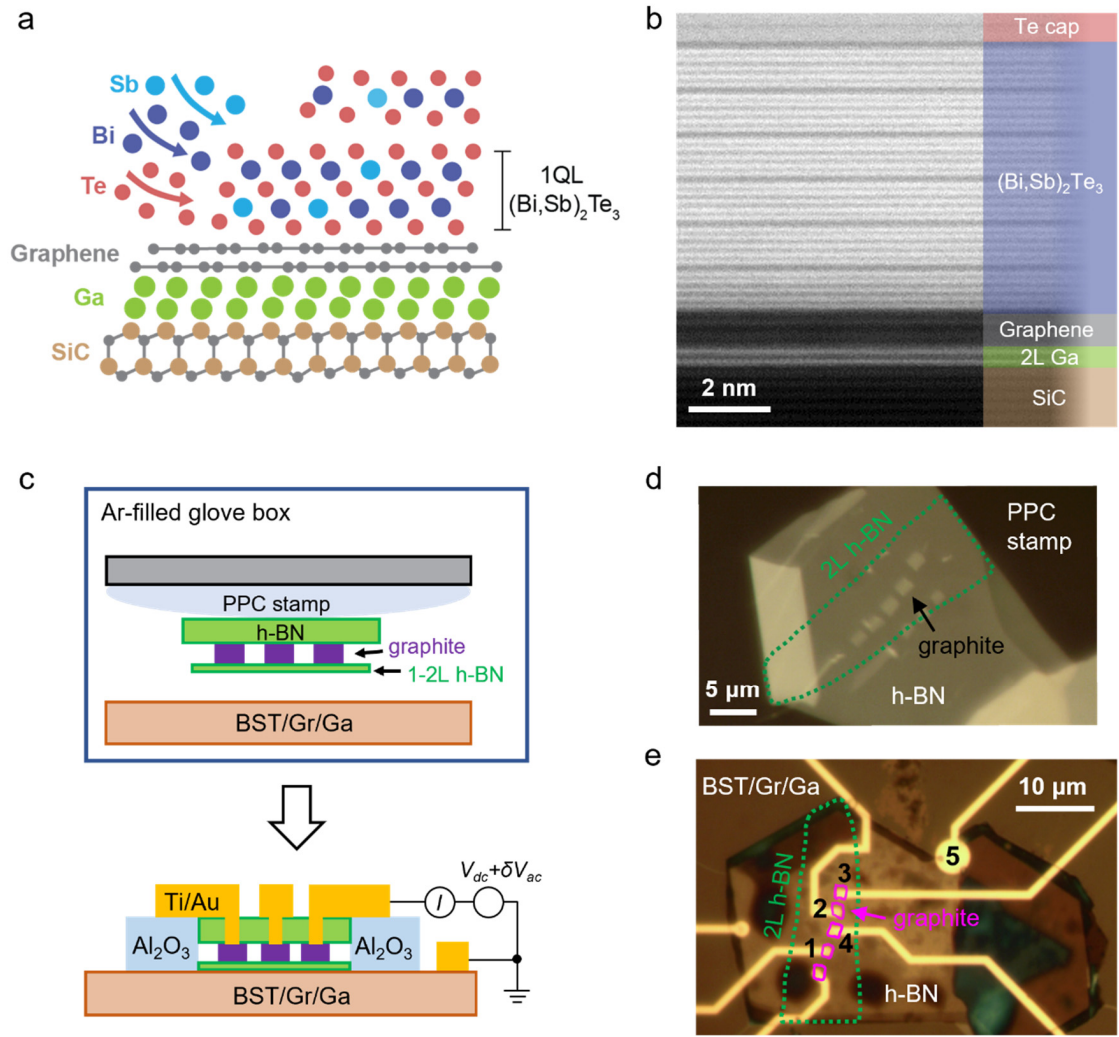

**Fig. 1 | Epitaxial growth of BST/Gr/Ga heterostructures and fabrication of tunnel junction devices. a,** Schematic of the growth combining CHet-growth of Gr/2L-Ga and MBE growth of $(Bi,Sb)_2Te_3$. **b,** A cross-sectional STEM image showing the atomically sharp hetero-interfaces between 6QL $(Bi,Sb)_2Te_3$, epi-graphene, and the 2L-Ga. **c,** Schematics of the van der Waals tunnel junction construction, device structure and the differential tunneling conductance measurement setup where a d.c. voltage ($V_{dc}$) with a small a.c. modulation ($\delta V_{ac}$) is applied to the graphite electrode. **d,** Optical image of an h-BN/graphite/2L h-BN stack on a poly(propylene) carbonate (PPC) stamp. The graphite squares are produced by atomic force microscope oxidation nanolithography[23] and are ~3 μm² in area. **e,** Optical image of device 02T-5QL with four tunnel junctions as labeled. Contacts 1-4 are made to the graphite electrodes and contact 5 directly contacts the BST/Gr/Ga film.

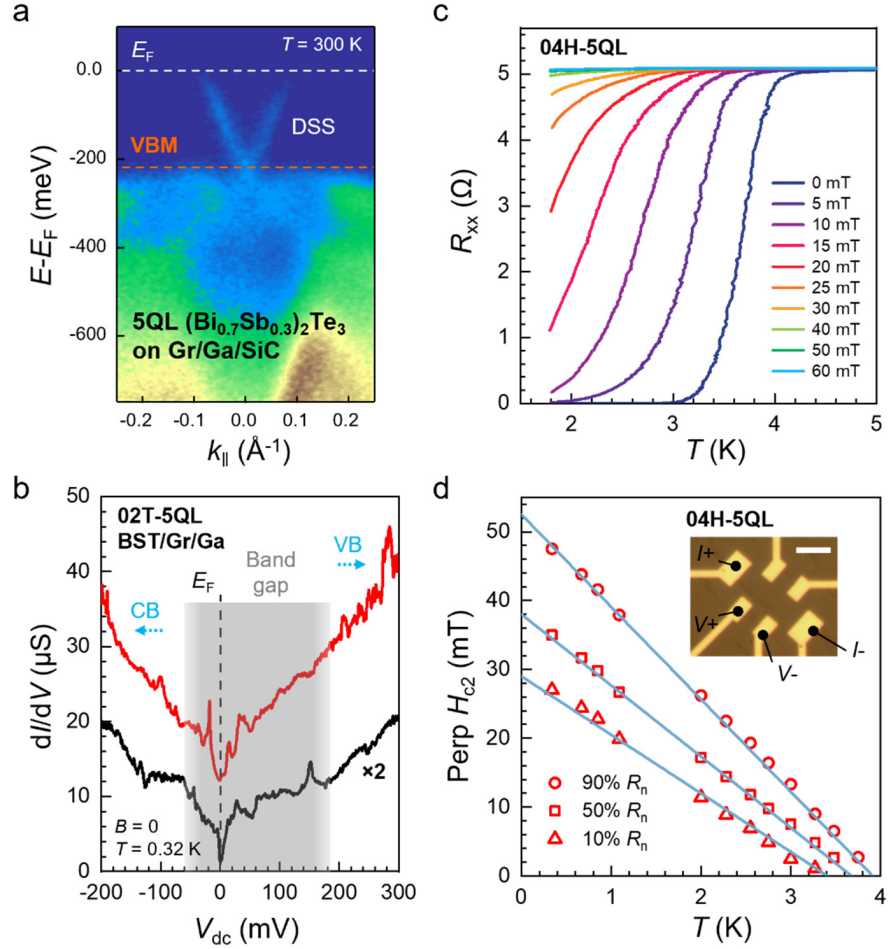

**Fig. 2 | Coexistence of Dirac surface states and superconductivity in BST/Gr/Ga heterostructures. a,** Room-temperature ARPES band map of a 5QL $(Bi_{0.7}Sb_{0.3})_2Te_3$/Gr/Ga film. $k_{\parallel}$, in-plane momentum. DSS, Dirac surface states; VBM, valence band maximum. **b,** $dI/dV$ versus $V_{dc}$ obtained on 02T-5QL made on the same film. Red and black curves are from different junctions. The red curve is shifted vertically by 10 µS for clarity. A positive d.c. bias probes states below the Fermi level. The grey shade box corresponds to the bulk band gap region of the BST film, consistent with the ARPES measurement in **a**. CB and VB represent the locations of the conduction band and the valence band, respectively. **c,** $T$-dependent $R_{xx}$ obtained on 04H-5QL at selected magnetic fields. The field is applied perpendicular to the film. **d,** Out-of-plane upper critical field $H_{c2}$ as a function of $T$. Solid lines are linear fits using the 2D Ginzburg-Landau model. Values of $H_{c2}$ are read from the measurements shown in Fig. S3f of the SI at fields where $R_{xx}$ reaches 90%, 50% and 10% of the normal state resistance $R_n$. The inset shows the optical image of 04H-5QL, where Ti/Au electrodes are directly deposited onto the film. Current source ($I+$), drain ($I-$), and high/low voltage ($V+/-$) probes are labelled. Scale bar, 5 µm.

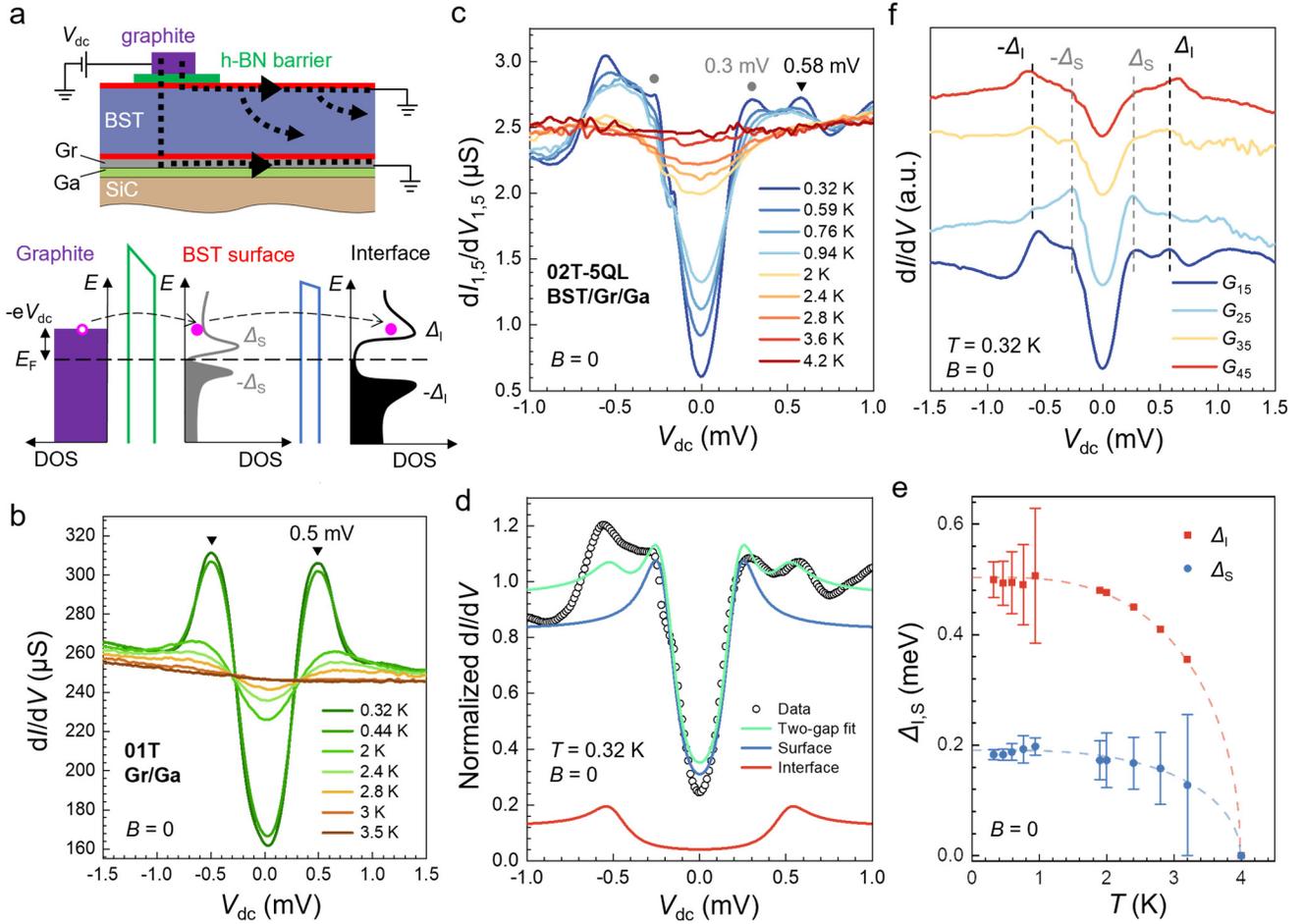

**Fig. 3 | Proximity-induced superconductivity in BST/Gr/Ga heterostructures. a,** Schematic of the two-gap tunneling process and the corresponding density of states diagrams. **b,** $dI/dV(V_{dc})$ obtained on device 01T (Gr/Ga without the growth of BST) at selected temperatures. **c,** $dI/dV(V_{dc})$ obtained on 02T-5QL between contacts 1 and 5 at selected temperatures. We examine the magnetic field dependence of the spectra in Fig. S5b to exclude coincidental features from the background, such as the sharp peak at $V_{dc} = -0.6$ mV. **d,** Fits to the 0.32 K data in **c** using a two-gap BTK model showing the contributions from the BST surface tunneling (blue trace), and the BST/Gr/Ga interface tunneling (red trace), and the total (green trace). The spectra weight of the blue trace is 88%. Only data at positive d.c. biases are fitted. **e,** $\Delta_S$ and $\Delta_I$ versus temperature obtained from the best fits to the BTK model. The dashed lines plot the BCS temperature dependence of a SC gap given in Eq. (1) with a common $T_c$ of 4 K. At $T \gtrsim 1.9$ K, we fix $\Delta_I$ following the red dashed line to obtain more accurate fitting of $\Delta_S$. Error bars represent 95% confidence interval of the fit. **f,** $dI/dV(V_{dc})$ obtained on four different junctions $G_{15}$-$G_{45}$, where the two subscript numbers represent the contacts between which the conductance is measured, on 02T-5QL. Curves are normalized by their normal state conductance and stacked vertically for clarity. All show features associated with two SC gaps. See Figs. S4-6 of the SI for the full BTK analysis on Gr/Ga, 5QL and 10QL BST/Gr/Ga films.

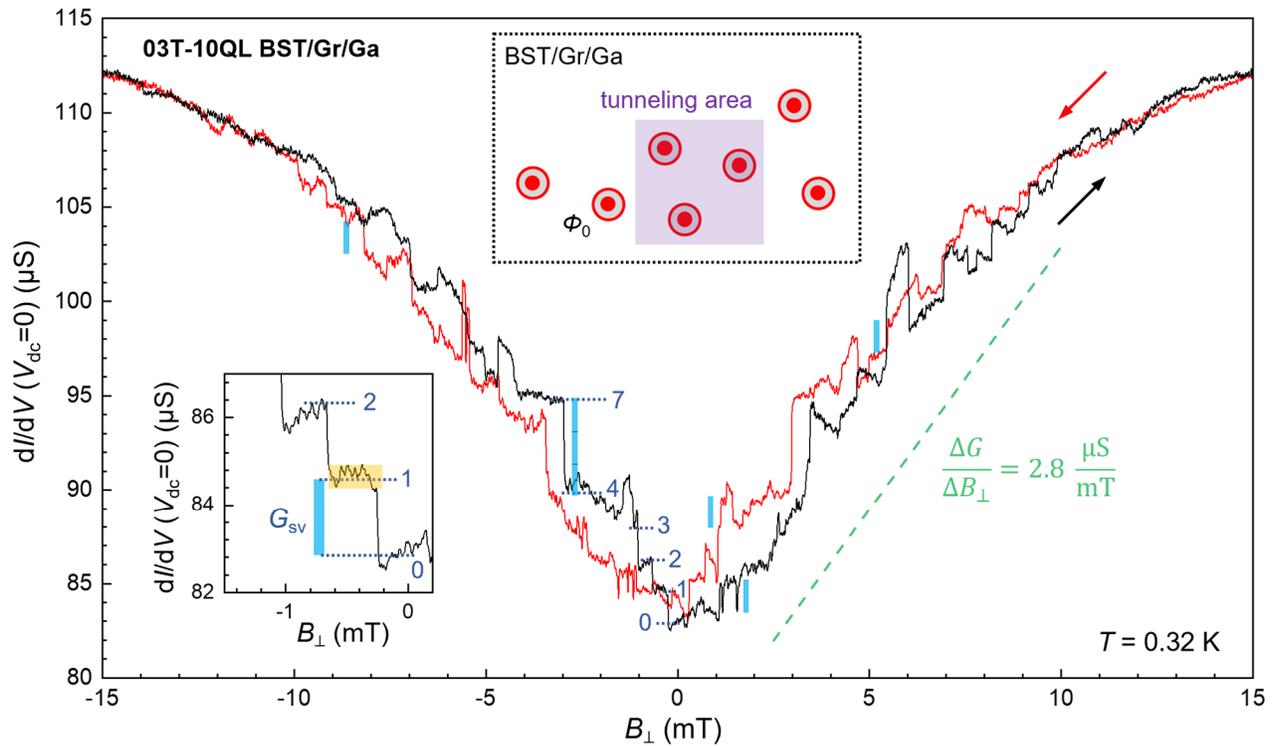

**Fig. 4 | Evidence of vortex trapping and single vortex signature in tunneling conductance.** Zero-bias $dI/dV$ measured on field upsweep (black trace) and downsweep (red trace) showing vortex-induced conductance change, down to the effect of a single vortex. Sweep rate = 10 µT/s. The lower inset expands on the black curve near $B_\perp = 0$. We attribute the smallest conductance step $G_{sv}$ (vertical light blue in the inset) to the effect of a single vortex $G_{sv} = 1.7$ µS. The horizontal orange bar marks the noise level in our measurement. Conductance plateaus corresponding to the first few vortices are marked by dashed lines and labeled in the plot. The upper inset shows a schematic of Abrikosov vortices penetrating the tunnel junction area defined by the size of the graphite contact $A \approx 3$ µm². The green dashed line is calculated from Eq. (2) using $A = 3.4$ µm² and agrees with data very well. Data from junction J2 of 03T-10QL (see descriptions in Section S5 of the SI).

**Methods**

**Confinement heteroepitaxy growth of 2D-Ga**

Gr/2D-Ga was grown using the CHet process described in Ref. [15]. Briefly, silicon carbide (II-VI Inc.) was diced into 1×0.5 cm or 1×1 cm substrates and pre-cleaned via a 20-minute soak in Nano-Strip (VWR, 90% sulfuric acid, 5% peroxymonosulfuric acid, <1% hydrogen peroxide). Subsequently, mono- or bi-layer epitaxial graphene was synthesized via silicon sublimation from the silicon carbide substrate for 20 minutes at 1800 °C in a 700 Torr argon atmosphere. Prior to intercalation, the graphene was subjected to an oxygen plasma (50 sccm He, 150 sccm $O_2$ at 500 mTorr and 50 W) in a M4L RF Gas Plasma system for 1 minute to generate defects that serve to facilitate metal diffusion to the graphene/SiC interface.

Intercalation was accomplished in a horizontal quartz tube (22 mm × 25 mm, inner × outer diameter) vacuum furnace, where gallium powder (Sigma Aldrich, 99.999% trace metals basis, 30-100 mg) was placed in an alumina crucible directly below a downward facing, plasma-treated Gr/SiC substrate. Prior to heating, the tube furnace was evacuated and backfilled with ultra-high purity (UHP) Ar. Finally, the sample and Ga powder were heated to 800 °C for 30 minutes under an UHP Ar or forming gas (3% $H_2$) environment at 300 – 700 Torr, with 50 sccm total gas flow. The sample was then cooled to room temperature.

**MBE growth of $(Bi,Sb)_2Te_3$**

$(Bi,Sb)_2Te_3$ films were grown in a commercial MBE system (Omicron Lab 10) with a base vacuum better than $2 \times 10^{-10}$ mbar. Graphene/Ga/SiC substrates were outgassed at ~350 °C for an hour prior to the growth of $(Bi,Sb)_2Te_3$. High purity Bi(99.9999%), Sb(99.9999%), and Te(99.9999%) were evaporated from Knudsen effusion cells. The flux ratio Te per (Bi + Sb) was set to be greater than 10 to prevent Tellurium deficiency in the film. The substrate was maintained at ~230 °C during growth. The growth rate of the $(Bi,Sb)_2Te_3$ film was about 0.2 QL/min. Reflection high energy electron diffraction (RHEED) was used to monitor the growth. Grown films were annealed at ~230 °C for 30 minutes to improve the crystal quality before cooling down to room temperature. After *in-situ* ARPES measurements, grown films were transferred to an Argon-filled glovebox using a small vacuum chamber to preserve the pristine $(Bi,Sb)_2Te_3$ surface.

*In-situ* **ARPES**

ARPES measurements were performed in a chamber with a base vacuum of $\sim 5 \times 10^{-11}$ mbar. The MBE-grown $(Bi,Sb)_2Te_3$ films were transferred to the ARPES chamber without breaking the ultrahigh vacuum. The photoelectrons were excited by an unpolarized He-$I_\alpha$ light (~21.2 eV), and a Scientia R3000 analyzer was used for the ARPES measurements. The energy and angle resolutions are ~10 meV and ~0.2°, respectively. All ARPES measurements were performed at room temperature.

**Cross-sectional STEM**

Cross-sectional transmission electron microscopy (TEM) samples were prepared using FEI Helios 660 Nanolab and FEI Scios 2 focused ion beam (FIB) systems. A thick protective amorphous carbon layer was deposited over the region of interest, and then $Ga^+$ ions (30 kV then stepped down to 1 kV to avoid ion beam damage to the sample surface) were used in the FIB to make the samples electron transparent for TEM images. High-resolution scanning transmission electron microscopy (STEM) was performed at 300 kV on a dual spherical aberration-corrected FEI Titan³ G2 60-300 S/TEM using a probe convergence angle of 24 mrad. The STEM images were collected using a high-angle annular dark-field (HAADF) detector with a collection angle of 50-300 mrad. Energy dispersive X-ray spectroscopy (EDS) maps were acquired on an FEI Talos F200X S/TEM operating at 200 kV.

**Transport studies**

All measurements were performed in a He-3 cryostat with a base temperature of 320 mK. We measure four-probe resistances using standard low-frequency lock-in techniques with an excitation current of 100 nA. To perform differential conductance measurement, we apply a d.c. voltage with a small a.c. modulation of 10-50 µV to the graphite contacts and measure the a.c. current using a lock-in amplifier (Stanford Research SR860). To control the magnet field finely, we use a Keithley 2400 sourcemeter to output current to the superconducting solenoid.

**Data availability:** The data needed to reproduce the main text and supplementary information figures are available on Zenodo (https://doi.org/10.5281/zenodo.7485214).

**Code availability:** The codes used in theoretical simulations and calculations are available from the corresponding author upon reasonable request.

# Supplementary Information:

# Proximity-Induced Superconductivity in Epitaxial Topological Insulator/Graphene/Gallium Heterostructures

Cequn Li[1], Yi-Fan Zhao[1], Alexander Vera[2,3], Omri Lesser[4], Hemian Yi[1], Shalini Kumari[2,3], Zijie Yan[1], Chengye Dong[5], Timothy Bowen[2,3], Ke Wang[6], Haiying Wang[6], Jessica L. Thompson[7], Kenji Watanabe[8], Takashi Taniguchi[9], Danielle Reifsnyder Hickey[3,6,7], Yuval Oreg[4], Joshua A. Robinson[1,2,3,5,6,7], Cui-Zu Chang[1], and Jun Zhu[1,2,5*]

1. Department of Physics, The Pennsylvania State University, University Park, PA 16802, USA
2. Center for 2-Dimensional and Layered Materials, The Pennsylvania State University, University Park, PA 16802, USA
3. Department of Materials Science and Engineering, The Pennsylvania State University, University Park, PA 16802, USA
4. Department of Condensed Matter Physics, Weizmann Institute of Science, Rehovot 760001, Israel
5. 2-Dimensional Crystal Consortium, The Pennsylvania State University, University Park, PA 16802, USA.
6. Materials Research Institute, The Pennsylvania State University, University Park, PA 16802, USA
7. Department of Chemistry, The Pennsylvania State University, University Park, PA 16802, USA
8. Research Center for Functional Materials, National Institute for Materials Science, 1-1 Namiki, Tsukuba 305-0044, Japan
9. International Center for Materials Nanoarchitectonics, National Institute for Materials Science, 1-1 Namiki, Tsukuba 305-0044, Japan.

* Correspondence to: jzhu@phys.psu.edu (J. Zhu)

## Contents

## 1. Additional ARPES measurements

Figure S1 shows the ARPES band maps of 5-10QL $(Bi_{0.7}Sb_{0.3})_2Te_3$/Gr/Ga (BST/Gr/Ga) heterostructures. In each panel, the Fermi level resides in the Dirac surface states of the $(Bi_{0.7}Sb_{0.3})_2Te_3$ film. The Dirac point shifts from -200 meV to -170 meV as the film thickness increases from 5QL to 10QL.

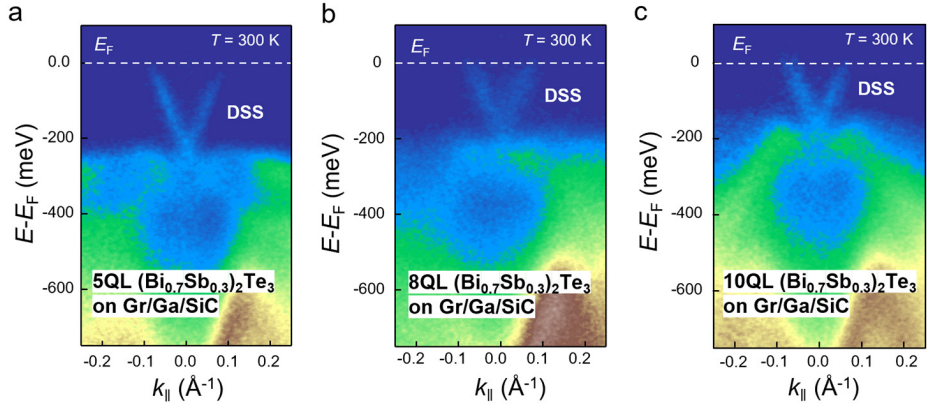

**Fig. S1 | Room-temperature ARPES band maps of 5-10QL $(Bi_{0.7}Sb_{0.3})_2Te_3$ grown on Gr/Ga.**

## 2. Fabrication of van der Waals tunnel junction

Figure S2 illustrates the steps we took to fabricate van der Waals (vdW) tunnel junctions on BST/Gr/Ga films. Grown films are transferred from the MBE chamber to an argon-filled glovebox using a vacuum suitcase with a base pressure of ~5×10$^{-8}$ mbar and kept in the glovebox to avoid oxidization and contamination of the BST surface. We first etch a long graphite stripe (3-6 nm in thickness) into squares of approximately 3 μm$^2$ in area using an atomic force microscope-based etching process[1] that avoids the use of polymer resists. These graphite squares serve as the top electrodes of the tunnel junction. We then construct an h-BN/graphite electrode/1-2L h-BN stack using vdW dry transfer techniques, a picture of which is shown in Fig. S2a. The stack is transferred to a BST/Gr/Ga film inside the glovebox. Next, we use standard lithographic techniques to deposit a 55-nm-thick Al$_2$O$_3$ film to surround the h-BN encapsulated area as shown in Fig. S2b. The Al$_2$O$_3$ film further protects the BST film from oxidization and also serves to isolate individual electrodes. We subsequently use e-beam lithography and established reactive ion etch recipe (CHF$_3$ 40 sccm/O$_2$ 4 sccm) to open circular windows with radius about 250 nm on the top h-BN sheet to expose the graphite electrodes (Fig. S2c). Finally, we pattern and deposit 5-nm Ti/60-nm Au metal leads to connect to the graphite electrodes. The optical image of a complete device is shown in Fig. S2d. Figure S2e shows the scanning electron microscope (SEM) image of another device.

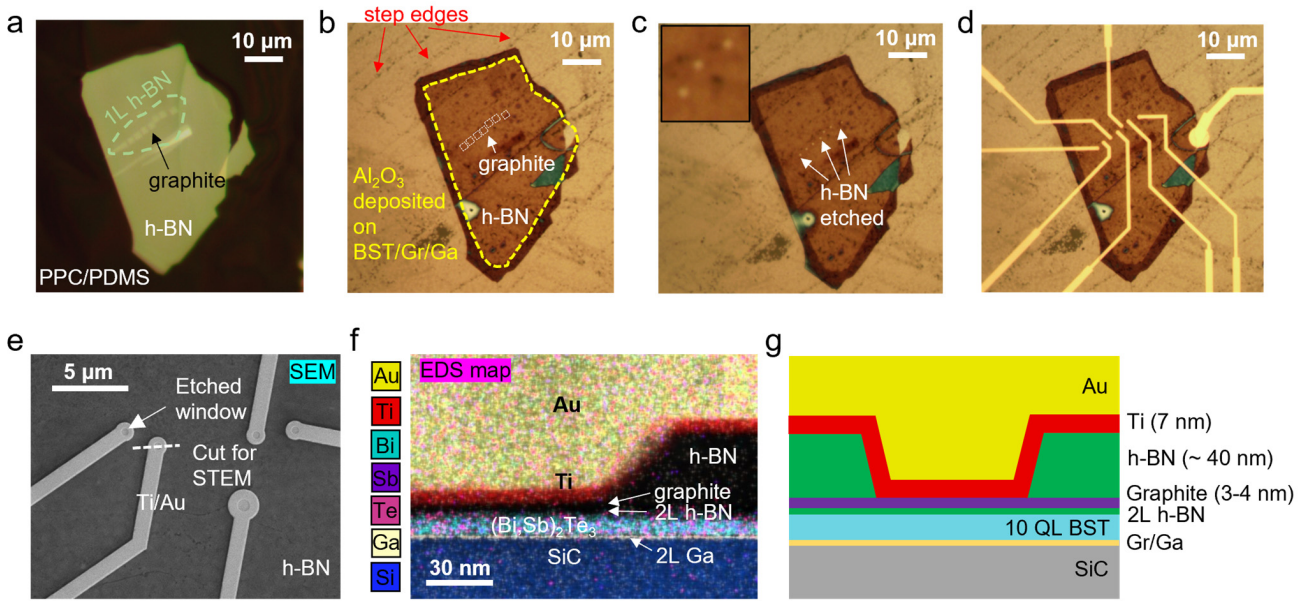

**Fig. S2 | Fabrication of van der Waals tunnel junction on BST/Gr/Ga. a,** A vdW stack of h-BN/graphite electrode/1L h-BN on a PPC/PDMS stamp. **b,** The stack transferred to the BST/Gr/Ga film and after the Al$_2$O$_3$ deposition. Outside the yellow dashed line, the BST/Gr/Ga film is covered with a 55nm-thick Al$_2$O$_3$ film. Red arrows point to step edges of the SiC substrate, where the intercalation of Ga may not be continuous. Tunnel junctions are placed within a single terrace. **c,** Circular windows of radius ~ 250 nm are etched into the top h-BN sheet to expose the graphite electrodes underneath. The inset shows a magnified image of a few windows. **d,** A completed device. **e,** SEM image of another device made on a 10 QL BST/Gr/Ga film. **f,** Cross-sectional EDS map of the tunnel junction. Light elements such as carbon, boron and nitrogen are not visible in this map. **g** shows a schematic diagram of the tunnel junction.

Focused ion beam (FIB) is used to cut and lift a slab along the dashed line in Fig. S2e. Scanning transmission electron microscopy (STEM) and elemental analysis are used to examine the cross sections. Figure S2f shows an energy dispersive x-ray spectroscopy (EDS) map of the van der Waals tunnel junction, with the corresponding layers labeled. This map agrees very well with the design of the structure schematically illustrated in Fig. S2g. The evaporated Au/Ti film makes contact to the graphite electrodes only.

3. **Four-terminal transport on BST/Gr/Ga**

We perform four-terminal transport on BST/Gr/Ga films using Hall bar devices shown in Fig. S3a. Au/Ti electrodes are evaporated to the surface of the film directly. Cross-sectional STEM images of the electrodes (Fig. S3b) show that at some locations, the metal film dips down to reach the Gr/Ga layer through voids of the BST film produced during its growth. Thus, these measurements, though conducted on the heterostructure film, reflects the superconducting properties of the 2D Ga. Fig. S3d-f show the temperature and magnetic field-dependent resistance in films of varying BST film thickness and Table S1 summarizes their growth parameters and superconducting characteristics. Overall we find the superconducting transition temperature $T_c$ of Ga varies between 3 to 4.2 K and the out-of-plane upper critical field $H_{c2}$ ranges 50–80 mT, consistent with previous findings of Briggs et al[2]. These measurements show that the growth of BST preserves the superconducting Ga layer very well and serve as reference to tunneling spectroscopy performed on the same films. Figure S3g plots the $V$-$I$ data of a 5QL BST/Gr/Ga film, where a critical current of 200 µA is observed. Although the current flow is limited to between the current leads, we can estimate the order of magnitude of the critical current density $J_c$ using the width of the quasi-hall bar $W = 4.7$ µm and a thickness of $d = 0.35$ nm for the 2L Ga film. $J_c$=0.14 A/µm$^2$ in this device, which is comparable to that of an elemental superconductor such as Al and Nb[3]. Measurements on other films show a typical $J_c$ of 0.01–0.1 A/µm$^2$.

| Film number | BST film thickness (QL) | x in (Bi$_x$Sb$_{1-x}$)$_2$Te$_3$ | $T_c$ (K) | Out-of-plane $H_{c2}$ (mT) |
|---|---|---|---|---|
| 01 | Gr/Ga | - | 3.1 | 79 |
| 02 | 5 | 0.7 | 4 | 80 |
| 03 | 10 | 0.7 | 4.1 | 70 |
| 04 | 5 | 0.4 | 3.9 | 47 |
| 05 | 8 | 0.7 | 4.2 | 62 |
| 06 | 10 | 0.7 | 3.5 | 73 |

**Table S1 | Thickness, stoichiometry and superconductivity parameters of BST/Gr/Ga films studied in this work.** We defined the critical temperature $T_c$ and the upper critical field $H_{c2}$ as the temperature and magnetic field at which the sample reaches 90% of its normal state resistance respectively.

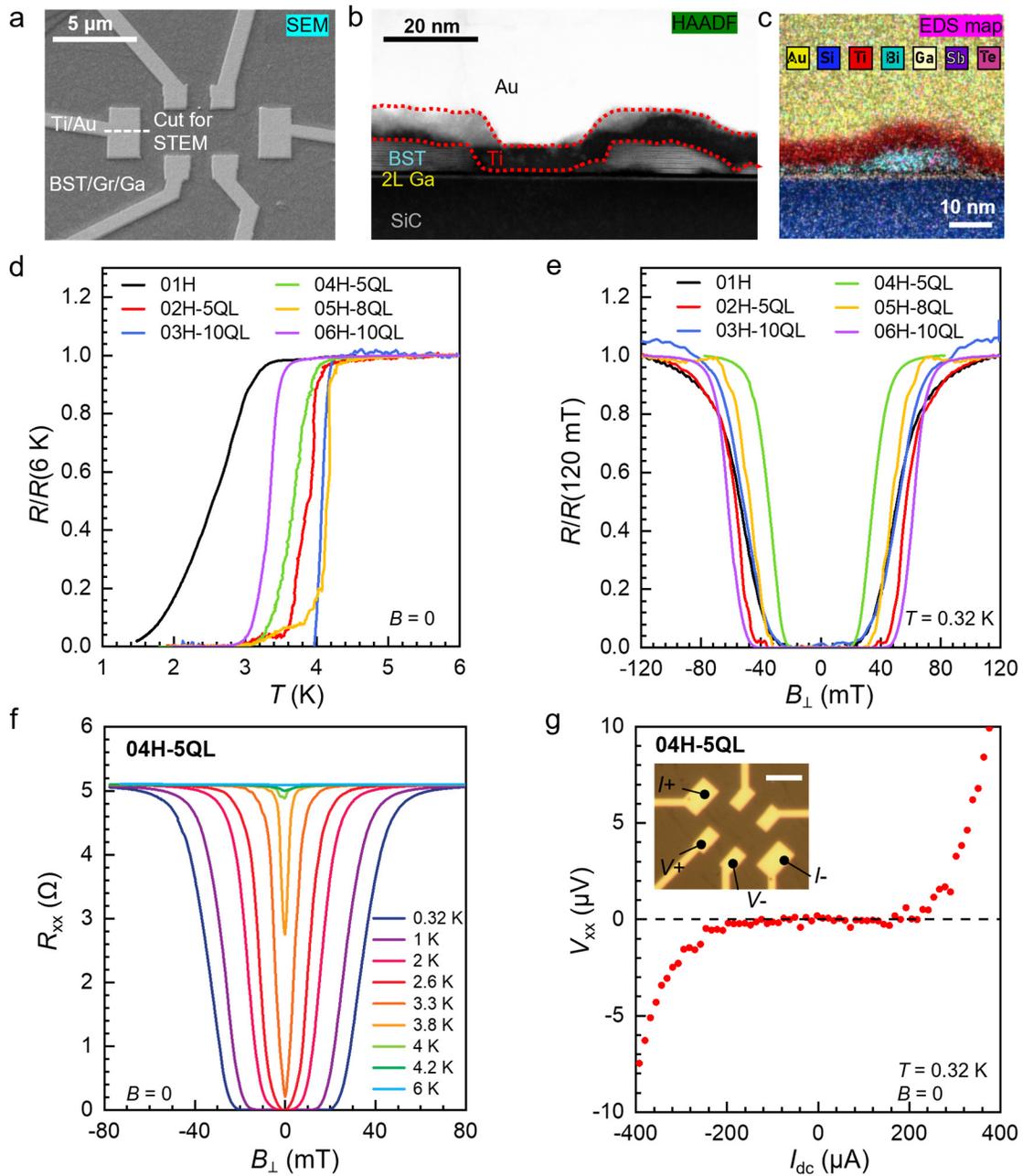

**Fig. S3 | Transport studies of BST/Gr/Ga heterostructures. a,** SEM image of 06H-10QL. **b,** Cross-sectional STEM image of a local area along the line cut shown in **a**. **c,** EDS map of a region close to **b** confirms the Ti layer outlined in red. It shorts to the Ga layer directly. **d,** Normalized temperature-dependent resistance $R(T)$ in devices as labeled. 01-06 correspond to the film number in Table S1. **e,** Normalized magnetic field-dependent resistance $R(B)$ of the same devices in **d**. **f,** $R(B)$ on device 04H-5QL at selected temperatures. **g,** $V_{xx}$ versus $I_{dc}$ on device 04H-5QL showing a large critical current of ~ 200 μA. The inset shows an optical image of the device. Metal electrodes are directly evaporated on the BST/Gr/Ga film to form a quasi-Hall bar structure. Scale bar: 5 μm.

## 4. Fitting the tunneling spectra on Gr/Ga using the Blonder-Tinkham-Klapwijk model

We fit the differential conductance $dI/dV$ vs $T$ to the Blonder-Tinkham-Klapwijk (BTK) model (Eqs. 1-3)[4], which contains three important parameters, i.e. the superconducting gap $\Delta$, the lifetime broadening $\Gamma$ of the Cooper pairs and the tunnel barrier strength $Z$. A completely transparent barrier has $Z = 0$.

$$\frac{dI}{dV} = G_N \int_{-\infty}^{\infty} dE [1 + A(E) - B(E)] \frac{\partial f(E - eV)}{\partial (eV)} \Lambda(E), \quad (1)$$

where

$$A(E) = \begin{cases} \dfrac{\Delta^2}{(E + i\Gamma)^2 + (1 + 2Z^2)^2 [\Delta^2 - (E + i\Gamma)^2]}, & E < \Delta \\ \dfrac{u_0^2 v_0^2}{\Upsilon^2}, & E \geq \Delta \end{cases} \quad (2)$$

and

$$B(E) = \begin{cases} 1 - A(E), & E < \Delta \\ \dfrac{(u_0^2 - v_0^2)^2 Z^2 (1 + Z^2)}{\Upsilon^2}, & E \geq \Delta \end{cases} \quad (3)$$

with $u_0^2 = \frac{1}{2}\mathrm{Re}(1 + \sqrt{\frac{(E+i\Gamma)^2-\Delta^2}{(E+i\Gamma)^2}})$, $v_0^2 = \mathrm{Re}(1 - u_0^2)$, and $\Upsilon^2 = [u_0^2 + Z^2(u_0^2 - v_0^2)]^2$. $f(E) = (1 + e^{E/k_B T})^{-1}$ is the Fermi-Dirac distribution function, and $\Lambda(E)$ is the empirical instrumental broadening term with the form of $\Lambda(E) = \sqrt{2\pi\sigma^2}\exp(-\frac{E^2}{2\sigma^2})$ where the variance $\sigma$ is chosen to be 2.5 times the ac excitation $\sigma = 2.5 V_{ac}$ following conventional practices. We have also produced fits with $\sigma = V_{ac}$ and $\sigma = 3V_{ac}$ and the changes to the fitting results are negligible. Eqs. 1-3 are implemented in MATLAB to fit data.

Figure S4 shows the full set of temperature- and magnetic field-dependent tunneling spectra acquired on device 01T (Fig. S4a-c), which is a Gr/Ga film without the BST growth, together with fits using the BTK model (Fig. S4d). The resulting $T$-dependent SC gap $\Delta(T)$, the tunnel barrier transparency $Z$ and the lifetime broadening $\Gamma$ of the Cooper pairs are given in Fig. S4e-f. $\Delta(T)$ is well described by the BCS theory, with a $T_c = 3.1$ K and $\Delta_0 = 0.395$ meV, consistent with four-terminal studies performed on the same film (See device 01H in Fig. S3). The tunnel barrier strength $Z$ slightly decreases with increasing temperature while the lifetime broadening $\Gamma$ increases somewhat with increasing $T$. These are consistent with naïve expectations. For instance, interaction with phonon modes at elevated temperatures may increase $\Gamma$ while thermally assisted tunneling results in an apparent increase in the transparency of the tunnel barrier. Fitting parameters obtained here serve as reference and consistency check to analyses of the BST/Gr/Ga junctions.

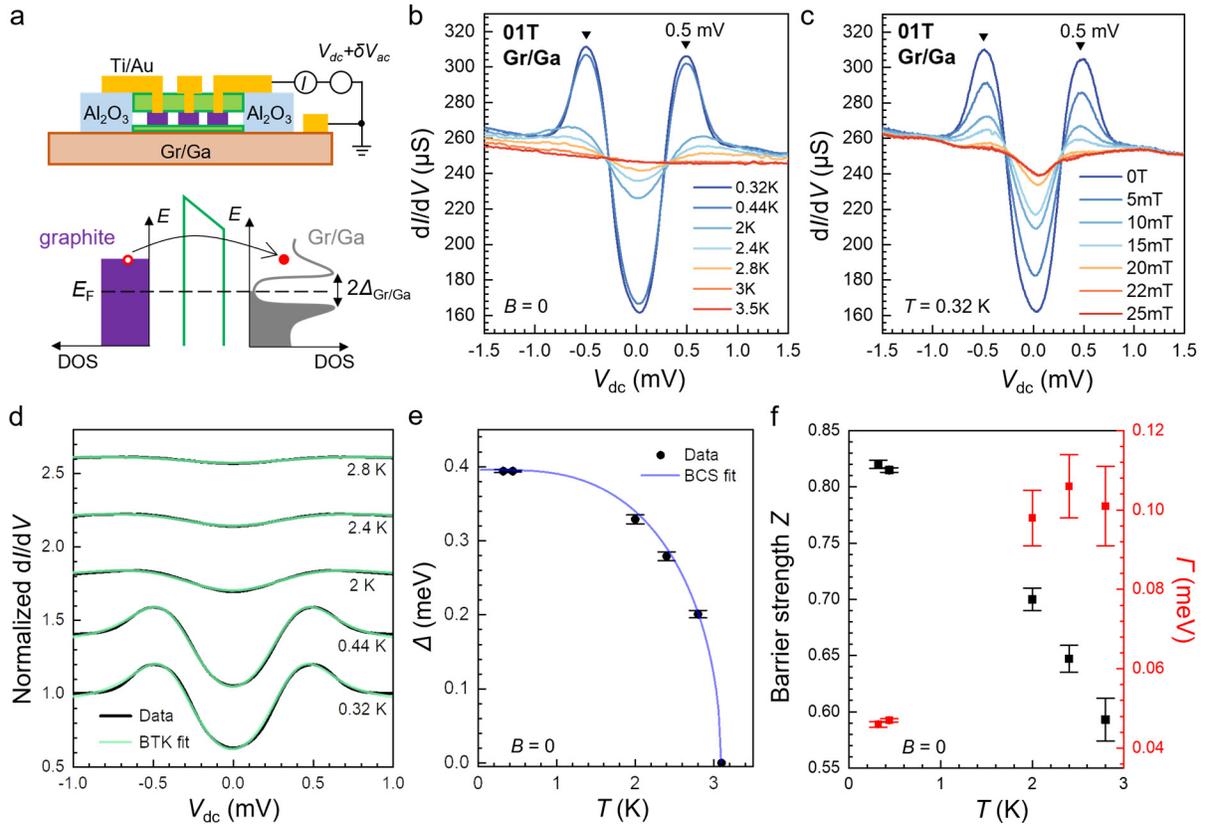

**Fig. S4 | Fitting the superconducting tunneling spectra on Gr/Ga using the BTK model. a**, Schematics of the tunnel junction and the tunneling process. **b, c,** $dI/dV$ versus $V_{dc}$ on device 01T at selected temperatures (**b**), and perpendicular magnetic fields (**c**). **d**, BTK fits at different temperatures. The fits capture data very well. **e**, $\Delta(T)$ obtained from fits in **d**. The blue solid line plots the BCS description of $\Delta(T) = \Delta_0 \tanh\left(1.74\sqrt{\frac{T_c}{T}-1}\right)$, where $T_c = 3.1$ K and $\Delta_0 = 0.395$ meV. **f**, The tunnel barrier strength parameter $Z$ and lifetime broadening parameter $\Gamma$ as a function of temperature obtained from the fits in **d**. Error bars represent 95% confidence interval of the fits.

## 5. Two-gap BTK fits on the 5QL and 10QL BST/Gr/Ga heterostructures

A two-gap BTK model[5] is used to fit the tunneling spectra in 5 and 10 QL TI/Gr/Ga heterostructures, where characteristics of two superconducting gaps are observed (Figs. S5 and S6). We consider the total tunneling current to be the sum of two contributions that originate from tunneling into the BST surface $(dI/dV)_S$ and the BST/Gr/Ga interface $(dI/dV)_I$ respectively:

$$(dI/dV)_{\text{total}} = f_S \times (dI/dV)_S + f_I \times (dI/dV)_I,$$

(4)

where $f_{S,I}$ represents the spectra weight of each contribution and $f_S + f_I = 1$. The two-gap fit contains 5 parameters, i.e, $f_S$, $\Delta_S$, $\Delta_I$, $\Gamma_S$, $\Gamma_I$ and a common $Z$ as it is mostly determined by the h-BN tunnel barrier.

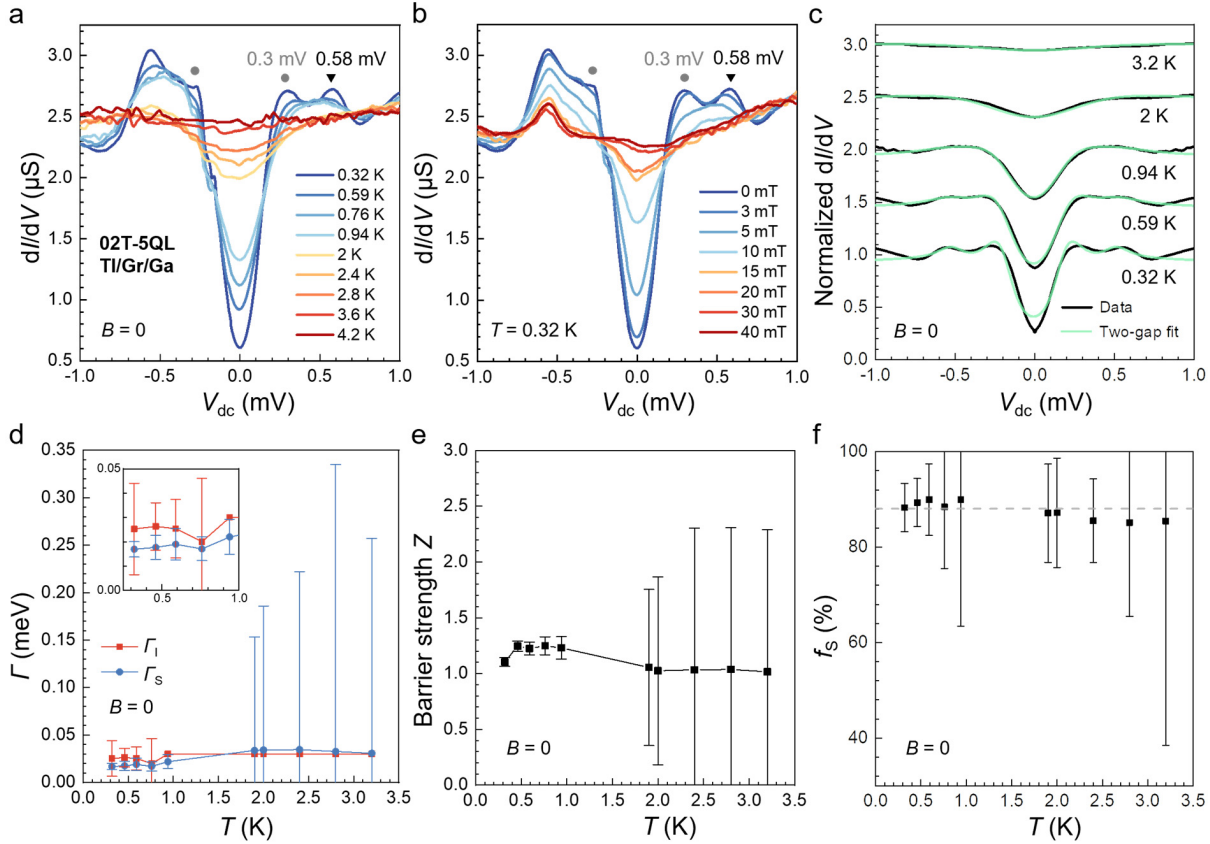

**Fig. S5 | Tunneling spectra on device 02T-5QL and two-gap BTK fits. a, b,** $dI/dV$ versus $V_{dc}$ at selected temperatures (**a**) and perpendicular magnetic fields (**b**). **c,** Two-gap BTK fits to the data in **a** at selected temperatures. As the $B$-dependent spectra in **b** show, certain features on the negative bias side may not be related to superconductivity. We have therefore chosen to only fit data on the positive bias side in **c**. **d,** Lifetime broadening parameters $\Gamma_S$ and $\Gamma_I$ as a function of temperature obtained from the fits in **c**. The low temperature value of $\Gamma_I \sim 0.03$ meV agrees with the results in Fig. S4 very well. To reduce fitting uncertainty, we have fixed $\Gamma_I \sim 0.03$ meV for $T \sim 1$ K and above, where $k_B T \gg \Gamma$ and thermal broadening dominates. This allows us to obtain fits for $\Gamma_S$, which comes out to be similar to $\Gamma_I$ although with large uncertainties. Part of this uncertainty originates from the anti-correlation between $\Gamma_S$ and the barrier strength $Z$. A large $Z$ reduces the zero-bias conductance, which requires a smaller $\Gamma_S$ to compensate. **e,** The barrier strength $Z$ as a function of temperature obtained from the fits in **c**. **f,** $f_S$ as a function of temperature obtained from the fits in **c**, showing that on average, tunneling into the TI surface contributes 88% of the spectra weight. Error bars represent 95% confidence interval of the fits.

Fig. S5 show the $T$-dependent and $B$-dependent tunneling spectra taken on device 02T-5QL, the BTK fits, and the parameters obtained from the fits. The main results, i.e. $\Delta_S(T)$ and $\Delta_I(T)$, are shown in Fig. 3 of the main text. Figs. S5d-f plot the remaining fitting parameters. The two-gap fits yield robust values of $\varDelta_S$ and $\varDelta_I$, especially at low temperatures, as their values are dictated by prominent features of the data such as the V shape near zero bias and the appearance of a second coherence peak. A barrier strength of $Z \sim 1$ is consistent among all the junctions we analyzed (Figs S4-6). The results for the lifetime broadening parameters are less reliable at $T > 1$ K where features become broadened and independent determinations become difficult. Nonetheless, they are consistent with results obtained on Gr/Ga junctions (Fig. S4) and on a 10QL BST/Gr/Ga heterostructure (Fig. S6). Our fits show that the measured $dI/dV$ is dominated by the tunneling into the TI surface states, with $f_S \sim 0.9$. This is also reasonable, given that an electron reaching the Gr/Ga interface needs to traverse the semi-insulating $(Bi_{0.7}Sb_{0.3})_2Te_3$ bulk.

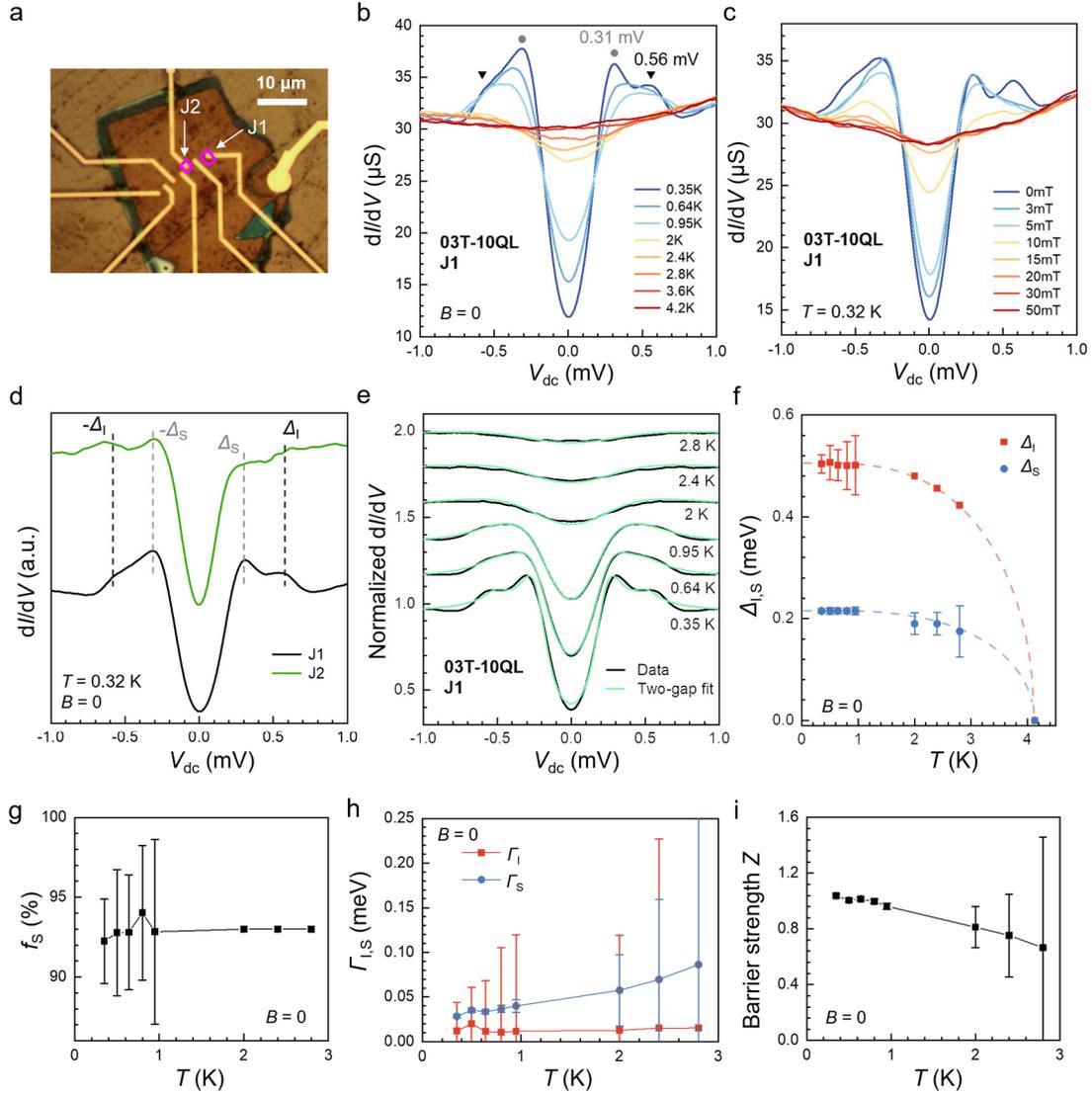

**Fig. S6 | Tunneling spectra on device 03T-10QL and two-gap BTK fits. a,** Optical image of 03T-10QL made on a 10QL (Bi$_{0.7}$Sb$_{0.3}$)/Gr/Ga film. J1 and J2 are two independent junctions. **b, c,** d$I$/d$V$ versus $V_{dc}$ at selected temperatures (**b**) and perpendicular magnetic fields (**c**) from J1. **d** plots the $T = 0.32$ K d$I$/d$V$ traces from both junctions J1 and J2. Curves are normalized by their normal state conductance and stacked vertically for clarity. **e,** Two-gap BTK fits to the data in **a** at selected temperatures. Only data at positive biases are analyzed. Data at negative biases are replica of the positive side. **f,** $\Delta_S(T)$ and $\Delta_I(T)$ obtained from the fits in **e**. The red and blue dashed lines represent fits to the temperature dependence of a BCS gap with a common $T_c = 4.1$ K, similar to Fig. 3e of the main text. **g,** $f_S$ as a function of temperature obtained from the fits in **e** showing that below 1 K, 93% of the tunneling current goes into the TI surface. Above 2 K, $f_S$ is fixed at 0.93 to enable more accurate fits for other parameters. **h,** Lifetime broadening parameters $\Gamma_S$ and $\Gamma_I$ as a function of temperature obtained from the fits in **e**. Values obtained here are consistent with those found in Figs. S4 and S5. **i,** Barrier strength $Z$ as a function of temperature obtained from the fits in **e**, which is also consistent with other junctions studied. Error bars represent 95% confidence interval of the fits.

Similar tunneling spectroscopy measurements and analysis are performed on two other T-5QL junctions made on film 02 and two T-10QL junctions made on film 03. Features associated with two SC gaps are consistently observed in all of our devices, confirming that a proximity-induced gap opening in the surface states of the BST film is a robust finding of our experiment. The measurements and analysis of 03T-10QL are shown in Fig. S6.

**Table S2** summarizes the proximity-induced $\Delta_S$ and the primary superconducting gap $\Delta_I$ obtained in different tunnel junctions made on BST films of different thicknesses $d$. The thickness dependence of $\Delta_S$ is not evident, likely due to the local variations of the film thickness under each junction. A study of $\Delta_S(d)$ will require a more precise control of the film thickness.

| BST film thickness (QL) | $T_c$ (K) | Junction | $\Delta_I$ (meV) | $\Delta_S$ (meV) | $\Delta_S/\Delta_I$ (%) |
|---|---|---|---|---|---|
| 0 | 3.1 | 01T | 0.395±0.003 | - | - |
| 5 | 4 | $G_{15}$ of 02T-5QL | 0.5±0.03 | 0.19±0.01 | 37±4 |
|   |   | $G_{25}$ of 02T-5QL | 0.49±0.05 | 0.2±0.003 | 41±4 |
|   |   | $G_{45}$ of 02T-5QL | 0.51±0.03 | 0.23±0.01 | 46±5 |
| 10 | 4.1 | J1 of 03T-10QL | 0.5±0.02 | 0.215±0.004 | 42±2 |
|   |   | J2 of 03T-10QL | 0.5±0.03 | 0.17±0.01 | 34±4 |

**Table S2 | $\Delta_I$, $\Delta_S$, and the ratio $\Delta_S/\Delta_I$ found in different devices.** $\Delta_I = \Delta_0$ in Gr/Ga.

## 6. Evidence of Abrikosov vortex probed by tunneling conductance

Discrete zero-bias conductance changes and hysteresis in $B$-field sweep due to the bunching and trapping of Abrikosov vortices are commonly observed in Gr/Ga and BST/Gr/Ga films we studied. Fig. S7a, c provide additional examples in devices 01T Gr/Ga and 02T-5QL BST/Gr/Ga. We identify the unit of conductance change $G_{sv}$, which all other conductance jumps are multiple integers of, and attribute it to the entry/exit of a single vortex. This enables us to calculate the number of vortices $N_{\text{vortex}}$ in the tunnel junction using

$$N_{\text{vortex}} = \frac{G(B) - G(0\text{T})}{G_{sv}},$$

(5)

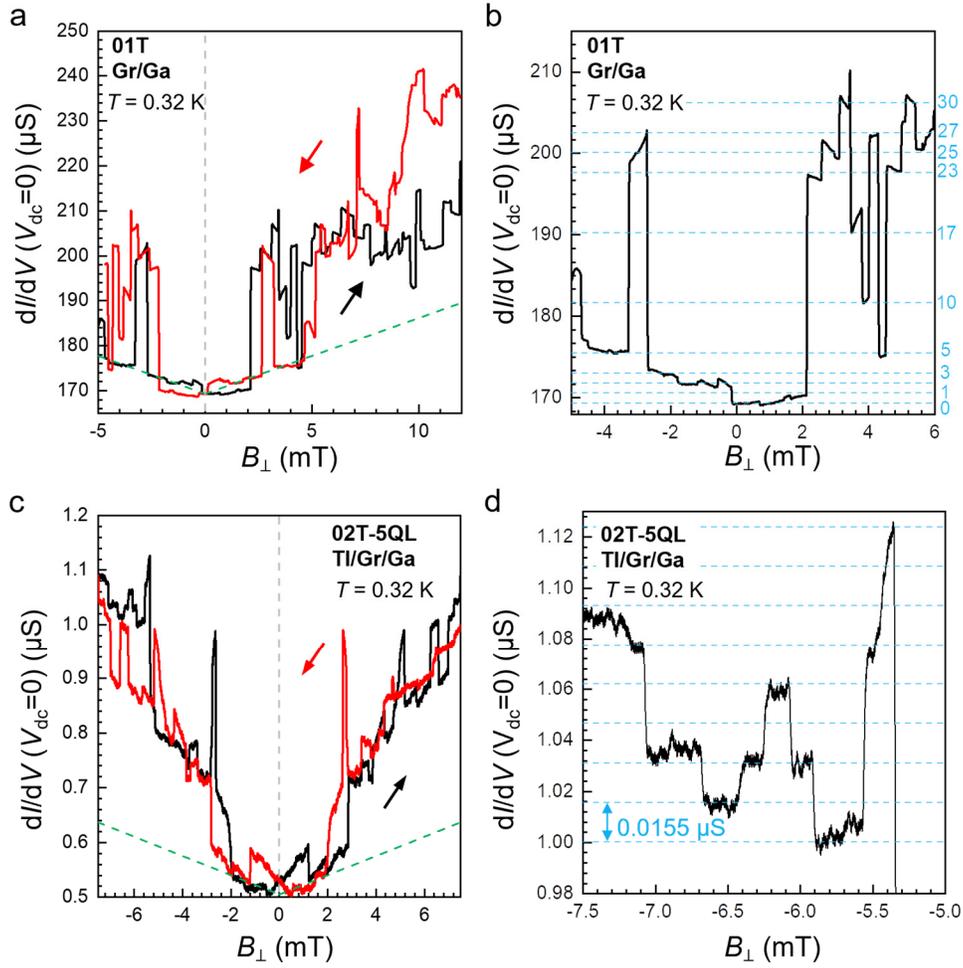

**Fig. S7 | Single vortex signature and vortex pinning in Gr/Ga and TI/Gr/Ga films. a**, Zero-bias $dI/dV$ versus $B_\perp$ in device 01 T (**a**) and 02T-5QL (**c**) showing hysteresis and discrete conductance jumps. The green dashed line is the calculated $\Delta G/\Delta B$ using Eq. (6) and the junction area $A = 2.82$ μm² in **a** and $A = 2.49$ μm² in **c**. The magnetic field is swept at a slow rate of 5–10 μT/s to allow for the establishment of steady state. $T = 0.32$ K. **b**, Conductance plateaus in **a** are labeled by $N_{vortex}$ given by Eq. (5). $G_{sv} = 1.23$ μS. **d**, Expanded data around $B_\perp = -6.5$ mT in **c** showing discrete conductance jumps in units of 1, 2, and 3 $G_{sv}$, where $G_{sv} = 0.0155$ μS corresponds to the minimum conductance jump due to a single vortex.

We label $N_{vortex}$ corresponding to individual conductance plateaus in device 01T in Fig. S7b. Figure S7d shows examples of 1, 2, and 3 vortex-induced conductance change in 02T-5QL. As $B$ increases, the number of vortices increases following $\Phi = B \times A = N_{vortex} \times \Phi_0$, where $A$ is the tunnel junction area given by the top graphite electrode and $\Phi_0 = 2.068$ mT·μm² is the magnetic flux quantum. Together with Eq. (5), we obtain the average rate of conductance change vs $B$-field:

$$\frac{\Delta G}{\Delta B} = \frac{A \times G_{sv}}{\Phi_0}.$$

(6)

Eq. (6) is plotted as a green dashed line in Figs. S7a and c for devices 01T and 02T-5QL respectively. They capture the trend of data around $B = 0$ quite well. At $B$ greater than a few mT, the conductance fluctuates both up and down with large amplitude, suggesting vortices move in and out of the tunnel junction area as bundles[6]. We expect that defects, such as domain walls in the film, to play an important role in the bunching and pinning of the vortices and details will vary from film to film and junction to junction. Indeed, Fig. S8 compares J1 and J2 made on the same film 03. J1 exhibits much larger conductance fluctuations than J2, suggesting stronger pinning effects.

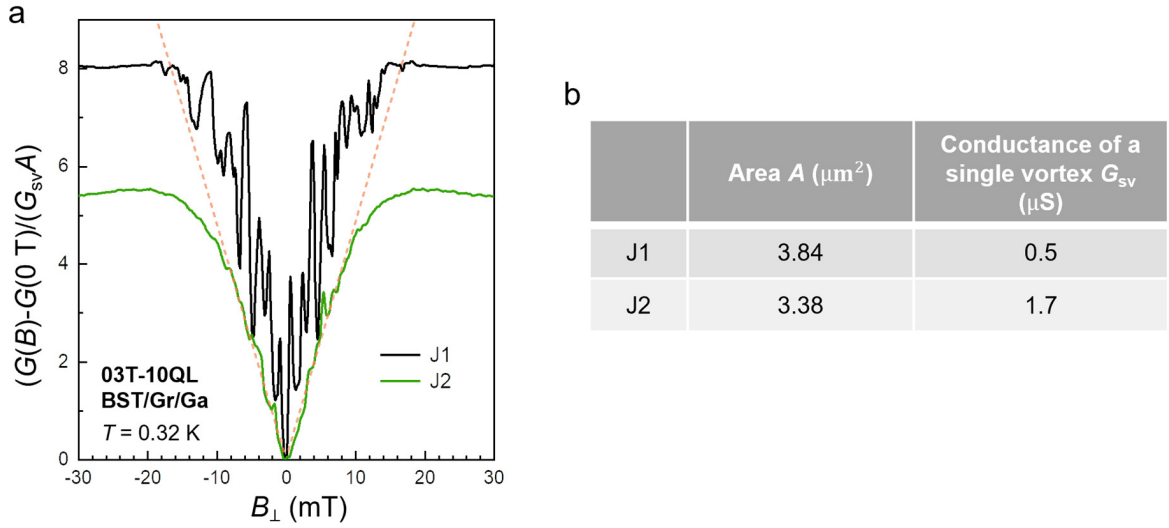

**Fig. S8 | Vortex pinning in different junctions made on 03T-10QL. a,** plots the vortex density $N_{\text{vortex}}/A$ calculated from $\frac{G(B)-G(0\,\text{T})}{G_{\text{sv}} \times A}$ for junctions J1 and J2, respectively. The orange dashed line plots $B/\Phi_0$. By definition, $N_{\text{vortex}}/A = B/\Phi_0$. J2 follows the orange dashed line very well, indicating smooth evolution of the Abrikosov vortex lattice while J1 exhibits large deviations from the expected rate, suggesting strong pinning effects. Field sweep rate: 100 µT/s. Parameters used in the analysis are given in **b**. Data of J2 are shown in Fig. 4 of the main text.